# Structural, electrical and magnetic studies of Co:SnO$_2$ and (Co,Mo):SnO$_2$ films prepared by pulsed laser deposition

S. Dalui[1,2,*], S. Rout[1], A.J. Silvestre[2], G. Lavareda[3], L.C.J. Pereira[4], P. Brogueira[5] and O. Conde[1]

[1]University of Lisbon, Physics Dept. and ICEMS, 1749-016 Lisboa, Portugal
[2]Instituto Superior de Engenharia de Lisboa and ICEMS, 1959-007 Lisboa, Portugal
[3]New University of Lisbon, Mater. Sci. Dept. and CTS, 2829-516 Caparica, Portugal
[4]Instituto Superior Técnico, ITN and CFMCUL, 2686-953 Sacavém, Portugal
[5]Instituto Superior Técnico, Physics Dept. and ICEMS, 1749-001 Lisboa, Portugal

**Abstract**

Here we report on the structural, optical, electrical and magnetic properties of Co-doped and (Co,Mo)-codoped SnO$_2$ thin films deposited on *r*-cut sapphire substrates by pulsed laser deposition. Substrate temperature during deposition was kept at 500 ºC. X-ray diffraction analysis showed that the undoped and doped films are crystalline with predominant orientation along the [101] direction regardless of the doping concentration and doping element. Optical studies revealed that the presence of Mo reverts the blue shift trend observed for the Co-doped films. For the Co and Mo doping concentrations studied, the incorporation of Mo did not contribute to increase the conductivity of the films or to enhance the ferromagnetic order of the Co-doped films.

[*] Corresponding author: ssdalui@fc.ul.pt (Saikat Dalui)

## 1. Introduction

Recent trend and opportunity in the field of diluted magnetic semiconductors (DMS) based spintronic devices boosted the research on transition metal (TM) doped wide band gap semiconductors like $TiO_2$, $SnO_2$, ZnO and $In_2O_3$ [1-4]. TM-doped $SnO_2$ is particularly interesting in this field for its added benefits, e.g. transparency and chemical sensitivity, and has recently been investigated by several research groups [5-11]. Room temperature ferromagnetism (RTFM) was observed in Co-doped $SnO_2$ by Ogale *et al.* [10] for thin films prepared by pulsed laser deposition (PLD) as well as by Punnoose *et al.* [5] for chemically synthesized powders. Coey *et al.* [11] also reported on RTFM for Fe-doped $SnO_2$ films. Recently, Nomura *et al.* [12] reported the enhancement of the saturation magnetization of Fe-doped $SnO_2$ by codoping with $Sb^{5+}$, suggesting that the increase in carrier density induces ferromagnetism. A similar approach was taken previously by Behan *et al.* [13] who codoped PLD ZnO films with Mn and Al.

This work intends to explore the potential of Mo in (Co,Mo)-codoped $SnO_2$ films for obtaining a DMS with improved properties. We report on the structural, optical, electrical and magnetic properties of undoped and codoped thin films prepared by PLD.

## 2. Experimental

Undoped, Co-doped and (Co,Mo)-codoped $SnO_2$ thin films were grown on *r*-cut sapphire substrates by PLD. A KrF excimer laser of wavelength 248 nm and pulse duration 30 ns, operating at a repetition rate of 10 Hz, was used for the ablation of targets with the following nominal composition: (A) $SnO_2$, (B) $Sn_{0.995}Co_{0.005}O_2$, (C) $Sn_{0.97}Co_{0.03}O_2$, and (D) $Sn_{0.96}Co_{0.01}Mo_{0.03}O_2$. Prior to any experiment, the PLD chamber was evacuated to a base pressure lower than $1\times10^{-6}$ mbar. The laser beam was focused on the rotating targets at an incidence angle of 45° and the laser fluence at the target surface was 2 J cm$^{-2}$. Processing parameters such as target-to-substrate distance, $d_{ts}$, substrate temperature, $T_s$, and working pressure, $P$, were varied in order to achieve the growth of crystalline and uniform films displaying ferromagnetic behavior. Here we show the results obtained with the following parameters: $d_{ts} = 6$ cm, $T_s = 500$ °C, $P = 7 \times 10^{-4}$ mbar using Argon as background gas.

The crystallographic structure of the as-grown samples was analyzed by X-Ray Diffraction (XRD) while their microstructure was observed by Field Emission Scanning Electron Microscopy (FE-SEM) and Atomic Force Microscopy (AFM) with the images taken in tapping mode. The root mean square (RMS) surface roughness was evaluated over an area of 1μm × 1μm on AFM images. The thickness of the samples was measured on cross-section SEM micrographs by image analysis. The obtained values were in the range 340 ± 20 nm. Optical studies were performed by measuring transmittance in the wavelength region 200 – 800 nm using a spectrophotometer with a bare substrate in the reference beam path. The spectra were recorded with a resolution of 0.5 nm. Electrical measurements were done in the temperature range 283 – 373 K by two-point probe technique using aluminum contacts deposited by evaporation with an appropriate mask. The measured resistance values were in the kΩ range, therefore they should not be substantially affected by the contact resistances which are of the order of tens of ohms. The magnetic measurements were carried out at 300 K and 4 K using a superconducting quantum interference device (SQUID) magnetometer with a magnetic field applied parallel to film surface. Table 1 summarizes the main properties of the films.

## 3. Results and discussion

*3.1. Structural and microstructural analyses*

Fig. 1 shows the XRD patterns of the (A) undoped, (B) 0.5% Co-doped, (C) 3% Co-doped and (D) (1%Co, 3%Mo)-codoped $SnO_2$ thin films. The films are crystalline with dominant peaks at $2\theta = 33.89°$ and $2\theta = 71.26°$ assigned, respectively, to the (101) and (202) planes of the rutile-type cassiterite phase of $SnO_2$ (JCPDS no. 41-1445). Although a small peak is showing on the samples C and D close to 30º, which is presumably due to a contribution from the substrate, there is no other perceptible trace of an impurity phase coming from the dopant ions such as metallic Co or Mo or any of their oxides. The intensity of the (101) peak is very low and the (202) peak is almost absent on pattern A as compared with the other three patterns. This shows that sample A is poorly crystallized. In contrast to the undoped film A, the 0.5% Co-doped sample B is highly oriented along the [101] direction. Sample C is clearly polycrystalline exhibiting other peaks at $2\theta = 57.82º$ and $2\theta = 64.72º$ assigned to the (002) and (112) planes of $SnO_2$, respectively, while sample D stays in between (B) and (C) i.e., it is less textured than (B) and not so randomly oriented as (C). In order to evaluate the films' mosaicity, the rocking curves (RC) of the (101) reflection were measured and their full-width at half-maximum (FWHM) determined as 2.22º (A), 1.10º (B), 2.76º (C) and 1.63º (D) confirming the qualitative interpretation afore given. For comparison purposes, the measured RC-FWHM of the $Al_2O_3$ (012) reflection was 0.28º. The average size of the crystallites was calculated from the broadening of the (101) peak by using the Scherrer's formula [14] as 28.9 nm (A, undoped), 29.2 nm (B), 23.6 nm (C) and 23.7 nm (D) (Table1).

Fig. 2 displays, on the left side, the SEM micrographs of the four typical films and, on the right, the corresponding 2D AFM images. As can be seen the films are uniform with a granular surface microstructure. Sample C shows the smoother surface morphology which is corroborated by both techniques (*cf.* the vertical scale on the AFM picture). The RMS surface roughness values are 9.3 nm for film A (undoped), 7.7 nm for film B (0.5% Co), 1.05 nm for film C (3% Co), and 8.8 nm for film D (1%Co, 3%Mo). The explanation for the very low roughness of film C might be linked with its polycrystalline microstructure. It is known that $SnO_2$ grows epitaxially on *r*-cut sapphire with the in-plane orientation relationship [15]: $SnO_2$ (101) [010] || $Al_2O_3$ ($\bar{1}012$) [$1\bar{2}10$]. However, in our case, the film is poorly crystallized which may arise from a high concentration of oxygen vacancies due to the growth conditions [16,17]. When a small (0.5%) amount of Co is added during film growth, assuming that the $Co^{2+}$ ions substitute for the $Sn^{4+}$ ones, the number of oxygen vacancies is reduced [18] and films grow highly oriented with respect to the substrate (sample B). But, because the ionic radius of $Co^{2+}$ with six-fold coordination number is higher than that of $Sn^{4+}$, respectively, 74.5 pm and 69 pm [19], the addition of a higher amount of Co, as in sample C, would prevent the epitaxial growth trend favouring the growth of polycrystalline films with no preferred orientation. Small grains randomly oriented will give rise to smoother films. Sample D, with a lower concentration of Co, 1%, reverts the observed trend for the microstructure evolution since Mo ions (4+, 5+ or 6+) are smaller than $Sn^{4+}$ [19].

*3.2. Optical properties*

The optical transmission spectra of the four representative films are plotted in Fig. 3 revealing that the Co-doped ones are mostly transparent with transmittance higher than 85 % in the visible range. In particular, interference fringes can be seen on the spectrum of sample C attesting the smoothness of its surface, in agreement with the SEM and AFM results. In contrast, the undoped and (Co,Mo)-codoped samples are less transparent with an average transmittance of about 55 % which can be understood by considering the higher surface roughness of both samples and their higher oxygen deficiencies producing more scattering centers [20]. Indeed, the inclusion of Mo into the $SnO_2$ matrix results in the substitution of higher oxidized ions ($Mo^{5+}$, $Mo^{6+}$) for $Sn^{4+}$ leading to an increase of oxygen vacancies and thus producing a sample with optical characteristics similar to the undoped sample.

The optical band gap energies, $E_g$, of the samples were calculated using the Tauc plots (Fig. 3, inset) and are listed in Table 1. There are some specific trends of optical band gap variation for the $SnO_2$ films with and without doping. Undoped $SnO_2$ samples show a band gap value of 3.85 eV, in agreement with reported values [21], whereas for Co-doped samples an $E_g \sim 4$ eV was measured. On the other hand, for the (Co,Mo)-codoped sample the band gap diminished down to ~ 3.3 eV. The information about the optical band gap energies for Co-doped $SnO_2$ thin films is scarce. Fitzgerald *et al.* [22] mentioned no significant band gap variation of the $SnO_2$ films with Co-doping, whereas in ref. [23] $E_g$ increased from 3.76 to 4.04 eV when the Co doping level increased. Nevertheless, there is a broad range of reported values for undoped and Co-doped $SnO_2$ nanoparticles [24-28]. Processing parameters such as growth temperature or annealing temperature, crystallite size and doping level strongly affect the optical band gap. Ahmed *et al.* [24] reported a decrease in $E_g$ from 4.1 to 3.8 eV with rising the sintering temperature from 300 ºC to 700 ºC. Bouaine *et al.* [26] observed a red shift of the $E_g$ edge from 3.68 to 3.45 eV when inserting 2% of Co in the $SnO_2$ matrix. Hays *et al.* [27] observed a rather unusual behavior with $E_g$ decreasing for up to 1% Co-doping and then increasing with further increase in Co content. On the other hand, Alanko *et al.* [28] reported a band gap energy of 3.2 eV for 10% Fe-doped $SnO_2$ nanoparticles which represents a red shift of ~0.9 eV in comparison with the undoped $SnO_2$ nanoparticles. Beyond structural modifications, the variations in band gap can also be due to other effects like quantum confinement and Burstein-Moss (BM) effects [29-33]. The quantum confinement effect can be ruled out since the crystallite size determined by XRD is much higher than the Bohr radius of 2.7 nm for $SnO_2$ [32]. The band gap broadening, $\Delta E_g$, corresponding to the Burstein-Moss effect in a degenerate semiconductor is expressed as [34]:

$$\Delta E_g = \left(\frac{\hbar^2}{2m_{vc}^*}\right)\left(3\pi^2 n_e\right)^{2/3} \qquad (1)$$

where $n_e$ is the charge carrier concentration, $\hbar$ is the reduced Planck's constant and $m_{vc}^*$ is the valence-conduction band reduced effective mass. Although targets with small nominal Co concentrations were used, it might happen that the effective doping is higher in the films [22]. With metallic doping the carrier density increases causing band gap widening. Indeed, from the electrical measurements (discussed later) it is evident that film resistivity was reduced with the Co-doping. Further co-doping with Mo reverts the observed trend and the band gap narrows substantially, probably due to the higher oxidation state of Mo as compared with Sn, and therefore to the increase of the electron-electron and electron-impurity scattering [33]. Equation (1) allows to roughly estimate a carrier concentration of about $6\times10^{19}$ cm$^{-3}$ for the Co-doped SnO$_2$ samples using $m^*_{vc} = 0.275\ m_0$ [31] where $m_0$ is the electron mass. This value compares well with that of $1.4\times10^{19}$ cm$^{-3}$ reported by Fitzgerald *et al.* [22] for Co-doped SnO$_2$ films grown by pulsed laser deposition.

*3.3. Electrical properties*

Conductivity measurements were performed in the temperature range 283 – 373 K, assuming that the films are uniform and doping ions are uniformly distributed in the samples. All the films show semiconductor *like* behavior with an increase in conductivity with metallic doping, irrespective of their morphology and microstructure. The room temperature conductivity, $\sigma_{RT}$, for the representative samples A, B, C and D are given in Table 1 which are well within the range as reported by different groups [18,22]. The $\sigma_{RT}$ of the 0.5%Co-doped SnO$_2$ film is four times that of undoped samples, whereas the 3% Co-doped and (1%Co, 3%Mo)-codoped SnO$_2$ films show conductivity improvement compared to undoped films but less than the 0.5%Co one. This is probably due to the fact that with increased Co and Co/Mo doping the number of dopant ions becomes too high, acting as scattering centers and reducing the carrier mobility. Moreover, the existence of oxygen vacancies may not only contribute towards the increase of the carrier density but often they act as localized scattering centers. Figure 4 shows the Arrhenius plot of $\sigma$ as a function of temperature, for the as-grown films. The estimated activation energies are 42, 11, 36 and 35 meV for films A, B, C and D, respectively. These values are consistent with data reported for oxygen vacancies in SnO$_2$ [35].

*3.4. Magnetic properties*

Magnetic measurements were carried out at temperatures of 300 K and 4 K. The undoped $SnO_2$ film showed diamagnetic behavior, as commonly reported [22]. Magnetization measurements performed on the 0.5% Co-doped (not shown) and (1%Co, 3%Mo)-codoped $SnO_2$ films (Fig. 5, top panel) show a weak ferromagnetic signal with practically no hysteresis. This behavior is typical of materials displaying superparamagnetism and might be explained by the low concentration of the magnetic transition metal. The most promising result was obtained for the film doped with 3% Co, which shows ferromagnetic behavior (Fig. 5, bottom panel). The magnetization data display well-defined hysteresis curves, with coercive fields of 74 and 125 Oe at RT and 4 K, respectively. It should be noted that these results seem to contradict published data for this system, where absence of the ferromagnetic order was observed for Co concentration higher than 1% [5, 27]. Taking into account the previously discussed electrical properties of the 3% Co-doped sample, the carrier mobility is expected to be low and, therefore, the ferromagnetic order observed in our sample might be explained within the scope of the bound magnetic polaron (BMP) theory by Coey *et al.* [36].

**4. Conclusions**

Optical, electrical and magnetic properties of crystalline Co-doped and (Co,Mo)- codoped $SnO_2$ thin films deposited on *r*-cut sapphire substrates were investigated and the results were correlated with the structural and microstructural properties and doping. While Co-doped samples display a blue shift that could be explained by the Burstein-Moss effect, codoping with Mo reverted this tendency most probably due to the higher oxidation state of Mo compared with Sn. For the Co and Mo doping concentrations here investigated, the presence of Mo did not contribute to increase the conductivity of the films nor to enhance the ferrromagnetic order of the Co-doped films. Further investigation by X-ray photoelectron spectroscopy is currently underway aiming at a clear identification of the oxidation states for both Co and Mo ions, their concentration ratio and distribution along film depth.

**Acknowledgements**

This work was supported by the Portuguese Foundation for Science and Technology (FCT), grant no. PTCD/CTM/101033/2008. S.D. acknowledges a Post-Doctoral grant funded by the aforementioned FCT research grant. S.R. acknowledges funding from FCT (grant no. SFRH/BPD/64390/2009).

**References**

[1] Y. Matsumoto, M. Murakami, T. Shono, T. Hasegawa, T. Fukumura, M. Kawasaki, P. Ahmet, T. Chikyow, S. Koshihara, H. Koinuma, Science 291 (2001) 854.

[2] Y. Tian, Y. Li, T. Wu, Appl. Phys. Lett. 99 (2011) 222503.

[3] H. Kim, M. Osofsky, M.M. Miller, S.B. Qadri, R.C.Y. Auyeung, A. Pique, Appl. Phys. Lett. 100 (2012) 032404.

[4] C. van Komen, A. Punnoose, M.S. Seehra, Solid State Comm. 149 (2009) 2257.

[5] A. Punnoose, J. Hays, V. Gopal, V. Shutthanandan, Appl. Phys. Lett. 85 (2004) 1559.

[6] K. Nomura, J. Okabayashi, K. Okamura, Y. Yamada, J. Appl. Phys. 110 (2011) 083901.

[7] S. Ghosh, M. Mandal, K. Mandal, J. Magn. Magn. Mater. 323 (2011) 1083.

[8] A.F. Lamrani, M. Belaiche, A. Benyoussef, A. ElKenz, E.H. Saidi, J. Magn. Magn. Mater. 323 (2011) 2982.

[9] Y. Xiao, S. Ge, L. Xi, Y. Zuo, X. Zhou, B. Zhang, L. Zhang, C. Li, X. Han, Z. Wen, Appl. Surf. Sci. 254 (2008) 7459.

[10] S.B. Ogale, R.J. Choudhary, J.P. Buban, S.E. Lofland, S.R. Shinde, S.N. Kale, V.N. Kulkarni, J. Higgins, C. Lanci, J.R. Simpson, N.D. Browning, S. Das Sarma, H.D. Drew, R.L. Greene, T. Venkatesan, Phys. Rev. Lett. 91 (2003) 077205.

[11] J.M.D. Coey, A.P. Douvails, C.B. Fitzgerald, M. Venkatesan, Appl. Phys. Lett. 84 (2004) 1332.

[12] K. Nomura, C.A. Barrero, K. Kuwano, Y. Yamada, T. Saito, E. Kuzmann, Hyperfine Interact. 191 (2009) 25.

[13] A.J. Behan, A. Mokhtari, H.J. Blythe, D. Score, X-H. Xu, J.R. Neal, A.M. Fox, G.A. Gehring, Phys. Rev. Lett. 100 (2008) 047206.

[14] B.D. Cullity, S.R. Stock, Elements of X-ray Diffraction, 3rd ed., Prentice Hall, N.J.,

2001.

[15] J.E. Dominguez, X.Q. Pan, L. Fu, P.A. Van Rompay, Z. Zhang, J.A. Nees, P.P. Pronko, J. Appl. Phys. 91 (2002) 1060.

[16] E.L. Boulbar, E. Millon, J. Mathias, C. Boulmer-Leborgne, M. Nistor, F. Gherendi, N. Sbai, J.B. Quoirin, Appl. Surf. Sci. 257 (2011) 5380.

[17] R. Perez-Casero, J. Perrière, A. Gutierrez-Llorente, D. Defourneau, E. Millon, W. Seiler, L. Soriano, Phys. Rev. B 75 (2007) 165317.

[18] D. Menzel, A. Awada, H. Dierke, J. Schoenes, F. Ludwig, M. Schilling, J. Appl. Phys. 103 (2008) 07D106.

[19] A. Kelly, K.M. Knowles, Crystallography and Crystal Defects, 2nd ed., J. Wiley & Sons, Ltd., 2012.

[20] S. Chacko, N.S. Philip, V.K. Vaidyan, Phys. Stat. Sol. (a) 204 (2007) 3305.

[21] R. Dolbec, M.A.El Khakani, A.M. Serventi, M. Trudeau, R.G. Saint-Jacques, Thin Solid Films 419 (2002) 230.

[22] C.B. Fitzgerald, M. Venkatesan, L.S. Dorneles, R. Gunning, P. Stamenov, J.M.D. Coey, P.A. Stampe, R.J. Kennedy, E.C. Moreira, U.S. Sias, Phys. Rev. B 74 (2006) 115307.

[23] M.-M. Bagheri-Mohagheghi, M. Shokooh-Saremi, Physica B 405 (2012) 4205.

[24] A.S. Ahmed, A. Azam, M. Shafeeq, M. Chaman, S. Tabassum, J. Phys. Chem. Solids 73 (2012) 943.

[25] L.Z. Liu, X.L. Wu, J.Q. Xu, T.H. Li, J.C. Shen, P.K. Chu, Appl. Phys. Lett. 100 (2012) 121903.

[26] A. Bouaine, N. Brihi, G. Schmerber, C. Ulhaq-Bouillet, S. Colis, A. Dinia, J. Phys. Chem. C 111 (2007) 2924.

[27] J. Hays, A. Punnoose, R. Baldner, M.H. Engelhard, J. Peloquin, K.M. Reddy, Phys. Rev. B 72 (2005) 075203.

[28] G.A. Alanko, A. Thurber, C.B. Hanna, A. Punnoose, J. Appl. Phys. 111 (2012) 07C321.

[29] R.K. Gupta, Z. Serbetc, F. Yakuphanoglu, J. Alloy. Comp. 515 (2012) 96.

[30] S. Shet, K.S. Ahn, Y. Yan, T. Deutsch, K.M. Chrustowski, J. Turner, M. Al–Jassim, N. Ravindra, J. Appl. Phys. 103 (2008) 073504.

[31] B. Nasr, S. Dasgupta, D. Wang, N. Mechau, R. Kruk, H. Hahn, J. Appl. Phys. 108 (2010) 103721.

[32] E.J.H. Lee, C. Ribeiro, T.R. Giraldi, E. Longo, E.R. Leite, Appl. Phys. Lett. 84 (2004) 1745.

[33] H.A. Mohamed, Intern. J. Phys. Sci. 7 (2012) 2102.

[34] I. Hamberg, C.G. Granqvist, K.F. Berggren, B.E. Sernelius, L. Engström, Phys. Rev. B 30 (1984) 3240.

[35] E.R. Viana, J.C. González, G.M. Ribeiro, A.G. de Oliveira, Phys. Status Solidi RRL 6 (2012) 262.

[36] J.M.D. Coey, M. Venkatesan, C.B. Fitzgerald, Nature Mater. 4 (2005) 173.

**TABLE CAPTION**

Table 1. Nominal chemical composition of the targets, and structural, optical and electrical data for the as-grown samples.

**FIGURE CAPTIONS**

Fig. 1 XRD patterns of the undoped (A), 0.5% and 3% Co-doped (B and C, respectively) and (1%Co, 3%Mo)-codoped (D) $SnO_2$ thin films. The S label denotes the substrate peaks.

Fig. 2 SEM micrographs (left) and corresponding 2D AFM images (right) for the undoped (A), 0.5% and 3% Co-doped (B and C, respectively) and (1%Co, 3%Mo)-codoped (D) $SnO_2$ thin films.

Fig. 3 Optical transmission spectra of the undoped and doped $SnO_2$ films. The inset shows Tauc plots for all the samples.

Fig. 4 Conductivity as a function of temperature (Arrhenius plot) for the as-grown samples. Corresponding activation energies are 42 meV (film A), 11 meV (film B), 36 meV (film C) and 35 meV (film D).

Fig. 5 Magnetic moment *vs.* applied magnetic field (*M-B*) curves for the (1%Co, 3%Mo)-doped $SnO_2$ film (top panel) and the 3% Co-doped $SnO_2$ film (bottom panel), at 300 and 4 K.

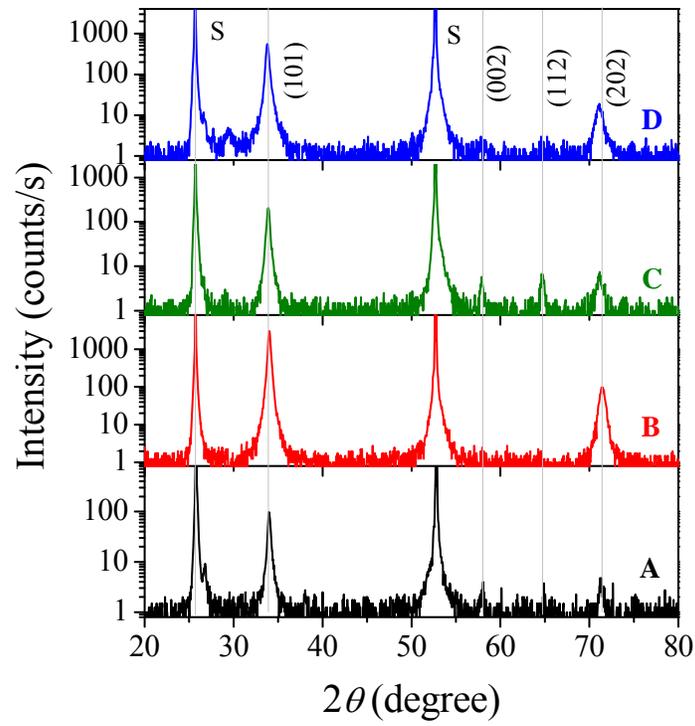

**Figure 1**

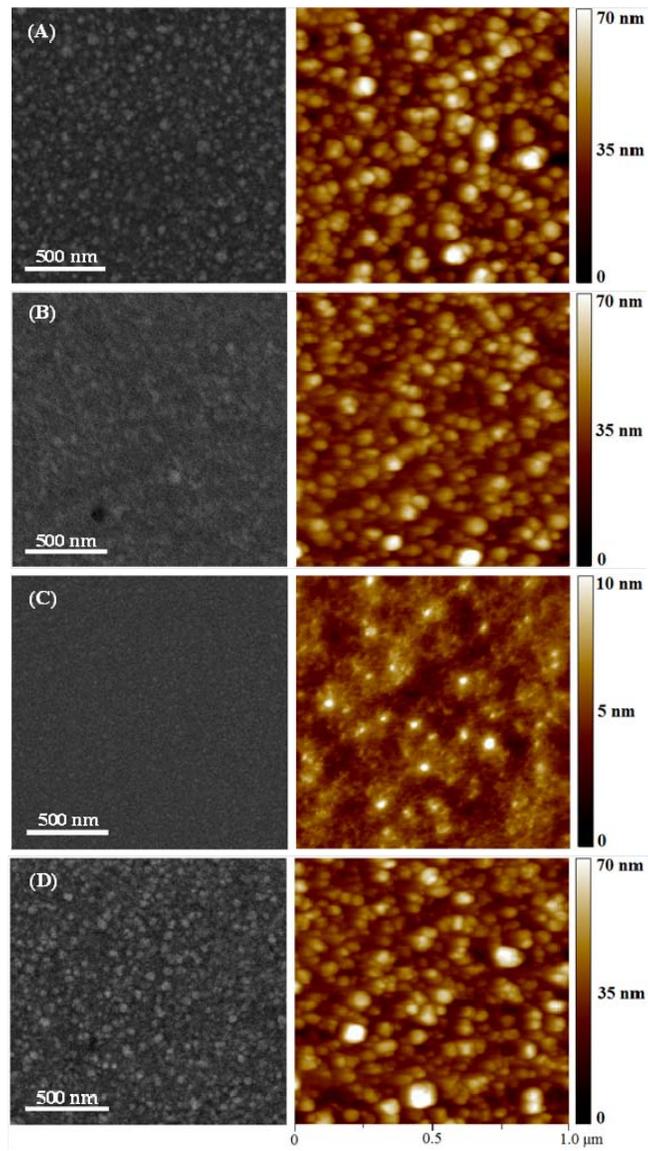

**Figure 2**

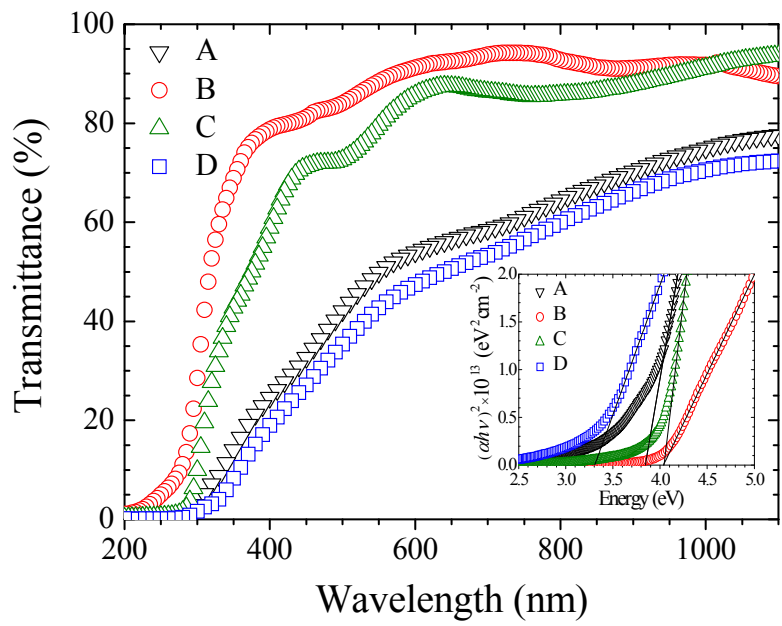

**Figure 3**

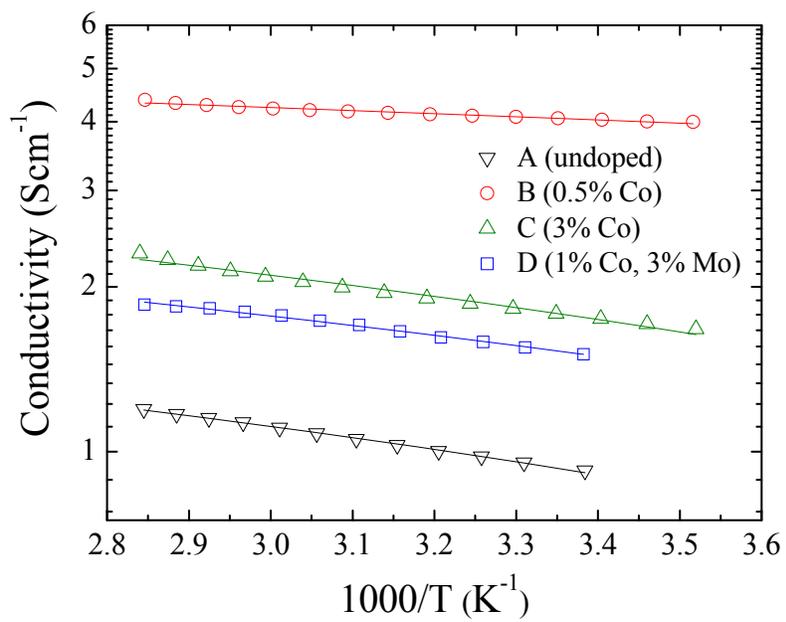

**Figure 4**

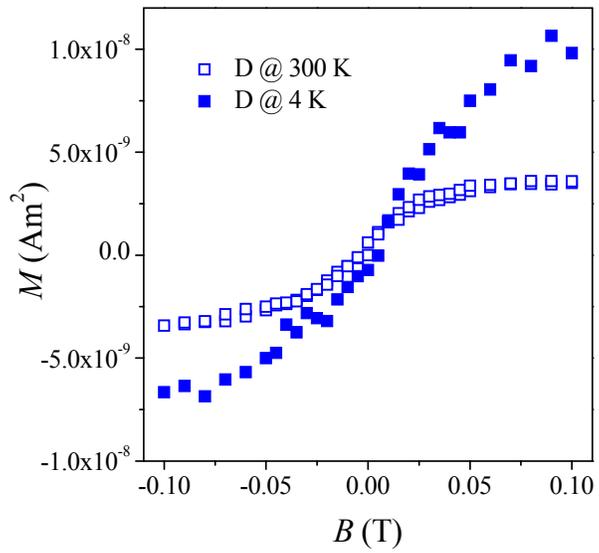
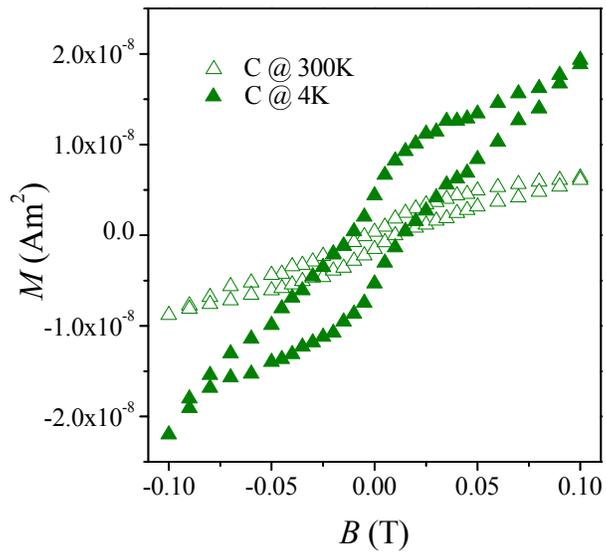

**Figure 5**

| Sample | Nominal composition of the targets | Thickness (nm) | Crystallite size (nm)[a] | RMS surface roughness (nm) | Optical band gap (eV) | RT conductivity (S cm$^{-1}$) |
|---|---|---|---|---|---|---|
| A | SnO$_2$ | 360 | 28.9 | 9.3 | 3.85 | 0.94 |
| B | 0.5%Co:SnO$_2$ | 339 | 29.2 | 7.7 | 4.04 | 4.06 |
| C | 3%Co:SnO$_2$ | 318 | 23.6 | 1.05 | 4.06 | 1.79 |
| D | (1%Co, 3%Mo):SnO$_2$ | 346 | 23.7 | 8.8 | 3.34 | 1.52 |

[a] Calculated from the (101) SnO$_2$ peak

**Table 1**